\documentclass[12pt,a4paper]{article}
\usepackage{latexsym,amsfonts}
\usepackage{exscale}
\usepackage{a4}
\usepackage[dvips]{graphicx} 
\usepackage{cite}

\newcommand{\beq}{\begin{equation}}
\newcommand{\eeq}{\end{equation}}
\newcommand{\beqa}{\begin{eqnarray}}
\newcommand{\eeqa}{\end{eqnarray}}

\newcommand{\pai}{\partial_{i}}
\newcommand{\paj}{\partial_{j}}
\newcommand{\pam}{\partial_{m}}
\newcommand{\paum}{\partial^{m}}
\newcommand{\pan}{\partial_{n}}

\newcommand{\da}{\partial_{\alpha}}

\newcommand{\daad}{\partial_{\alpha \dot \alpha}}

\newcommand{\Da}{D_{\alpha}}
\newcommand{\Db}{D_{\beta}}
\newcommand{\Dba}{\bar D_{\dot \alpha}}
\newcommand{\Dbb}{\bar D_{\dot \beta}}
\newcommand{\si}{\sigma_{i}}
\newcommand{\sj}{\sigma_{j}}
\newcommand{\sbi}{\bar \sigma_{i}}
\newcommand{\sbj}{\bar \sigma_{j}}
\newcommand{\sm}{\sigma^{m}}
\newcommand{\sn}{\sigma^{n}}
\newcommand{\sdm}{\sigma_{m}}

\newcommand{\sbm}{\bar \sigma^{m}}
\newcommand{\sbn}{\bar \sigma^{n}}
\newcommand{\sbdm}{\bar \sigma_{m}}
\newcommand{\sbdn}{\bar \sigma_{n}}
\newcommand{\smn}{\sigma^{mn}}

\newcommand{\sij}{\sigma^{ij}}

\newcommand{\te}{\theta}
\newcommand{\teb}{\bar{\theta}}
\newcommand{\teh}{\hat\theta}
\newcommand{\tehs}{\hat\theta^{'}}
\newcommand{\tehb}{\,\bar{\!\hat\theta}}
\newcommand{\tehbs}{\,\bar{\!\hat\theta^{'}}}

\newcommand{\dal}{\delta_{\alpha}}
\newcommand{\dl}{\delta_{\Lambda}}
\newcommand{\ds}{\delta_{\Sigma}}
\newcommand{\dlh}{\hat \delta_{\hat \Lambda}}
\newcommand{\La}{\Lambda}
\newcommand{\Lab}{\bar \Lambda}
\newcommand{\Lah}{\hat \Lambda}
\newcommand{\Labh}{\bar{\hat \Lambda}}
\newcommand{\Lahv}{\hat \Lambda (\Lambda,V)}

\newcommand{\Lap}{\Lambda^{\!'}}
\newcommand{\Lapp}{\Lambda^{\!''}}
\newcommand{\Labp}{\bar \Lambda^{\!'}}
\newcommand{\Lapv}{\Lambda^{\!'} ( \Lambda,V)}
\newcommand{\Labpv}{\bar \Lambda^{\!'} (\bar \Lambda,V)}
\newcommand{\Si}{\Sigma}

\newcommand{\Sihv}{\hat \Sigma (\Sigma,V)}

\newcommand{\Sipv}{\Sigma^{'} (\Sigma,V)}
\newcommand{\Sibpv}{\bar\Sigma^{'} (\bar\Sigma,V)}
\newcommand{\Fis}{\Phi^{'}_{+}}
\newcommand{\Fik}{\bar\Phi_{+}}

\newcommand{\Fihv}{\hat \Phi (\Phi,V)}

\newcommand{\Fipv}{\Phi^{'} (\Phi,V)}
\newcommand{\Fibpv}{\bar\Phi^{'} (\bar\Phi,V)}

\newcommand{\psib}{\bar \psi}
\newcommand{\vf}{A}
\newcommand{\vfs}{\bar A}
\newcommand{\la}{\lambda}
\newcommand{\lab}{\bar \lambda}

\newcommand{\al}{\alpha}
\newcommand{\alp}{\alpha'}

\newcommand{\vi}{v_{i}}
\newcommand{\vj}{v_{j}}
\newcommand{\vm}{v_{m}}

\newcommand{\vum}{v^{m}}

\newcommand{\tij}{\vartheta^{ij}}

\newcommand{\Paabb}{P^{\dot \alpha \alpha \dot \beta \beta}}
\newcommand{\qw}{\sqrt{2}}

\begin{document}

\thispagestyle{empty}

\rightline{LMU-TPW 2003-10}

\vspace{4em}
\begin{center}

{\Large{\textbf{Seiberg-Witten Map for Superfields on Canonically Deformed 
                $N\!=\!1$, $d\!=\!4$ Superspace}}}

\vskip 4em

{{\textbf{D\v zo Mikulovi\' c\footnote{dzo@theorie.physik.uni-muenchen.de}} }}

\vskip 2em

Universit\"at M\"unchen, Fakult\"at f\"ur Physik\\
Theresienstr.\ 37, D-80333 M\"unchen, Germany\\[1em]

\end{center}

\vspace{4em}

\begin{abstract}
\noindent
In this paper we construct Seiberg-Witten maps for superfields on canonically 
deformed $N\!=\!1$, $d\!=\!4$ Minkowski and Euclidean superspace. On Minkowski 
superspace we show that the Seiberg-Witten map is not compatible with locality,
(anti)chirality and supersymmetry at the same time. On Euclidean superspace we
show that there exists a local, chiral and supersymmetric Seiberg-Witten map 
for chiral superfields if we take the noncommutativity parameter to be 
selfdual, and a local, antichiral and supersymmetric Seiberg-Witten map for 
antichiral superfields if we take the noncommutativity parameter to be 
antiselfdual, respectively.
\end{abstract}
\newpage
\setcounter{page}{1}

\tableofcontents

\section{Introduction}
 
Field theories on canonically deformed space 
\beq
[\hat x^{i} ,\hat x^{j} ] = i \tij \, ,\qquad \tij=-\vartheta^{ji} \in 
\mathbb{R} \, ,
\eeq
have recently attracted much attention (for reviews and an exhaustive list of 
references see \cite{Konechny:0012,Douglas:0106,Szabo:0109}), mainly due to the
discovery of this noncommutative space in string theory 
\cite{Connes:1998,Douglas,Seiberg}. Based on the existence of different 
regularization procedures in string theory, Seiberg and Witten claimed in 
\cite{Seiberg} that certain noncommutative gauge theories are equivalent to 
commutative ones. In particular, they argued that there exists a map from a 
commutative gauge field to a noncommutative one, which is compatible with the 
gauge structure of each. This map has become known as the Seiberg-Witten map.

In \cite{Madore:0012,Jurco:0006,Jurco:0102,Jurco:0104} gauge theory on 
noncommutative space was formulated using the Seiberg-Witten map. In contrast 
to earlier approaches \cite{Armoni:0005,Bonora:0006,Bars:0103,Chaichian:0107},
this method works for arbitrary gauge groups. Using this method the problems 
of charge quantisation \cite{Hayakawa:99121,Hayakawa:99122} and tensor product 
of gauge groups \cite{Chaichian:0107} were solved and  the standard model and 
GUT's were formulated at the tree level on noncommutative space 
\cite{Calmet:0111,Aschieri:0205}.   

Non(anti)commutative superspaces naturally arise in string theory as well with
$x-x$ deformation (canonical deformation) \cite{Chu:1999}, $\te-\te$ 
deformation \cite{Ooguri:0302,Ooguri:0303,Seiberg:0305} and $x-\te$ deformation
\cite{deBoer:0302}. General deformed superspaces were first studied  more 
closely in \cite{Ferrara,Klemm} and recently in \cite{Ferrara:0307} and in 
connection with the supermatrix model in \cite{Park:0307}. Various aspects of 
field theory were considered mainly on the canonically deformed superspace and 
since the work of Seiberg 
\cite{Seiberg:0305} also on the $\te-\te$ deformed superspace with 
$N=\frac{1}{2}$ supersymmetry. On both spaces gauge theory was constructed 
without Seiberg-Witten map in \cite{Ferrara,Terashima:0002} and 
\cite{Seiberg:0305,Terashima:0306,Araki:0307}. Similar to the nonsupersymetric
case, without the Seiberg-Witten map only gauge theories for the $U(N)$ gauge 
group can be formulated on these spaces \cite{Terashima:0002,Terashima:0306}. 

On the canonically deformed superspace, gauge theory with Seiberg-Witten maps 
for component fields was considered in \cite{Putz:0205,Dayi:0309}. 
Seiberg-Witten map for superfields was briefly considered in 
\cite{Ferrara,Chekhov:2001}. In \cite{tocome} we will present the 
Seiberg-Witten map for superfields on the $\te-\te$ deformed superspace.

In this paper we will construct Seiberg-Witten map for superfields on 
canonically deformed $N\!=\!1$, $d\!=\!4$ superspace. First we will 
recapitulate some
well known properties of this space and construct the gauge theory coupled to
matter without and with Seiberg-Witten map for component fields. Thereafter
we will show that on Minkowski space the Seiberg-Witten map is not compatible 
with the requirements of locality, (anti)chirality and supersymmetry at the 
same time. We will 
present Seiberg-Witten maps for superfields in each case where one of these 
requirements is given up and construct the $U(1)$ gauge theory coupled to 
matter in terms of classical component fields using the nonsupersymmetric
Seiberg-Witten map. Finally we show that on Euclidean space there exists a 
local, chiral and supersymmetric Seiberg-Witten map for chiral fields
if $\tij$ is selfdual
\beq
\vartheta_{ij} = \frac{1}{2} \varepsilon_{ijmn} \vartheta^{mn} \, ,
\eeq
and a local, antichiral and supersymmetric one if the noncommutativity 
parameter is antiselfdual
\beq
\vartheta_{ij} = -\frac{1}{2} \varepsilon_{ijmn} \vartheta^{mn} \, .
\eeq
We use the conventions of \cite{WessBagger}.

\section{Canonically deformed superspace}

The canonically deformed real $N\!\!=\!\!1$, $d\!=\!4$ superspace, which we 
denote by $\hat\mathbb{R}^{4|4}$, has the following coordinate algebra 
\cite{Chu:1999}:
\beqa \label{eq2.1}
&& [\hat x^{i} ,\hat x^{j} ] ~=~ i \tij \, , \nonumber \\ \hat\mathcal{R}:
&&  [\hat x^{i} , \teh^{\alpha} ] ~=~ [\hat x^{i} , \tehb_{\dot\alpha} ] ~=~ 0
 \, , \\
&& \{ \teh_{\alpha} , \teh_{\beta} \} ~=~ \{ \tehb_{\dot \alpha} , \tehb_{\dot 
\beta} \} ~=~ \{ \teh_{\alpha} , \tehb_{\dot \alpha} \} ~=~ 0 \nonumber \, ,
\eeqa
with $\tij \!\in\!\mathbb{R}$. Noncommutativity is indicated by a hat. Since 
the Grassmann coordinates remain classical, in what follows "commutative" and 
"noncommutative" thus refers only to the bosonic part of coordinate algebra 
(\ref{eq2.1}). Noncommutative functions and fields are defined as elements of 
the noncommutative algebra 
\beq \label{eq2.2}
\hat \mathcal{A} = \frac{\mathbb{C} \big[ [\hat x^{i},\teh_{\alpha},
\tehb_{\dot\alpha}] \big]}{I_{\hat\mathcal{R}}} \, ,      
\eeq
where $I_{\hat\mathcal{R}}$ is the two-sided ideal created by the relations 
(\ref{eq2.1}). 

The derivatives can be introduced on a noncommutative space in a purely 
algebraic way \cite{Wess:1991}. In the case of canonical deformation, both the
bosonic and fermionic derivatives act on the coordinates as in the classical 
case
\beq \label{eq2.3}
[ \hat \pai , \hat x^{j} ] = \delta_{i}^{\,\,\,j} \, , \quad 
\{ \hat\partial_{\alpha},\teh^{\beta} \} = \delta_{\alpha}^{\,\,\,\beta} \, , 
\quad \ldots
\eeq
When $\tij$ is invertible, the bosonic derivatives are internal operations in
the algebra $\hat\mathcal{A}$ and they are given by    
\beq \label{eq2.7}
\hat \pai \hat F(\hat x,\teh,\tehb) = -i \vartheta_{ij}^{-1} [ \hat x^{j},\hat 
F(\hat x,\teh,\tehb) ] \, ,
\eeq
which leads to the classical relation
\beq \label{eq2.8}
[ \hat \pai , \hat \paj ] = 0 \, . 
\eeq 

It is convenient to use the star product formulation of the algebra. The star 
product on this noncommutative superspace is the well known Moyal-Wayl star 
product
\beqa 
\hat F(x,\te,\teb) * \hat G(x,\te,\teb) &=& e^{\frac{i}{2} 
\tij \frac{\partial}{\partial x^{i}} \frac{\partial}{\partial y^{j}}} \hat 
F(x,\te,\teb) \hat G(y,\te,\teb) |_{y \rightarrow x} \nonumber \\
&=& \hat F\hat G + \frac{i}{2} \,  \tij \pai \hat F \, \paj \hat G - 
\frac{1}{8} \tij \vartheta^{kl} \pai \partial_{k} \hat F  \paj \partial_{l} 
\hat G + \ldots \label{eq2.9}
\eeqa
Some useful properties of this product are:
\begin{itemize}
\item
involution
\beq
\overline{\hat F*\hat G} = \,\bar{\!\hat G} * \,\bar{\!\hat F} \, ,
\eeq
\item
Leibnitz rule 
\beq  \label{eq2.10}
\partial_{A} (\hat F*\hat G) = \partial_{A} \hat F* \hat G + \hat F * 
\partial_{A} \hat G \, ,
\eeq
where $\partial_{A}=(\partial_{m},\partial_{\alpha},
\bar\partial_{\dot\alpha})$,
\item
ordinary product of two functions under the integral\footnote{$d^{8}z=d^{4}x
d^{2}\te d^{2}\teb$.}
\beq
\int d^{8}z \, \hat F*\hat G = \int d^{8}z \,\hat F \hat G \, ,
\eeq 
if the functions vanish at infinity (which we demand),
\item
cyclicity of the product under the integral sign
\beqa 
\int d^{8}z\, \hat F*\hat G = \int d^{8}z\, \hat G*\hat F \, .  \label{eq2.11} 
\eeqa
\end{itemize}

\subsection{Symmetries}

\noindent
The algebra (\ref{eq2.1}) is covariant under the group of classical 
supertranslations parametrized by $(\hat a,\hat\xi,\!\bar{\,\hat\xi)}$   
\beqa \label{eq2.14}
\hat x'^{m} &=& \hat x^{m} + \hat a^{m} +i\teh\sm\,\bar{\!\hat\xi} - 
i\hat\xi\sm\tehb \, ,\nonumber \\
\tehs_{\alpha} &=& \teh_{\alpha} + \hat\xi_{\alpha} \, , \\
\tehbs_{\dot\alpha} &=& \tehb_{\dot\alpha} + \,\bar{\!\hat\xi}_{\dot\alpha} 
\, . \nonumber
\eeqa
which is generated by the complex charges $\hat Q_{\alpha}$, 
$\,\bar{\!\hat Q}_{\dot\alpha}$ and the four momentum $\hat P_{m}$. The 
generators have the same representation as in the classical case 
\beqa
\hat P_{m} &=& i \hat\pam \, , \label{eq2.15}\\
\hat Q_{\alpha} &=& \hat\partial_{\alpha} - i(\sm\tehb)_{\alpha} \hat\pam \ , 
\label{eq2.16} \\
\bar{\!\hat Q}_{\dot\alpha} &=& -\, \bar{\!\hat\partial}_{\dot\alpha} + 
i (\teh\sm)_{\dot\alpha} \hat\partial_{m} \, . \label{eq2.17} 
\eeqa
The supersymmetry algebra remains undeformed
\beqa
[ \hat P_{m},\hat P_{n} ] &=& 0 \, , \nonumber \\
\lbrack \hat P_{m},\hat Q_{\alpha} \rbrack &=&  \lbrack \hat P_{m}, 
\,\bar{\!\hat Q}_{\dot\alpha} \rbrack ~=~ 0 \, , \nonumber \\
\{ \hat Q_{\alpha},\hat Q_{\beta} \} &=& \{ \,\bar{\!\hat Q}_{\dot\alpha} ,
\,\bar{\!\hat Q}_{\dot\beta} \} ~=~ 0 \, , \nonumber \\
\{ \hat Q_{\alpha},\,\bar{\!\hat Q}_{\dot\beta} \} &=& 2 \, \sm_{\alpha\dot
\beta} \, \hat P_{m} \, , \label{eq2.18}
\eeqa 
because the algebra of derivatives is the same as in commutative case.

\subsection{Superfields}

\noindent
The covariant derivatives also have the standard form 
\beqa 
\hat{\,\Da} &=& \hat\da + i(\sm\tehb)_{\alpha} \hat\pam \, , \label{eq2.19} \\
{\bar{\!\hat D}}_{\dot\alpha} &=& - \!\bar{\,\hat\partial_{\dot\alpha}} - 
i(\teh\sm)_{\dot\alpha} \hat\pam \, , \label{eq2.20}
\eeqa
and we can define chiral superfields by $\!\bar{\,\hat D_{\dot\alpha}} 
\hat\Phi\!=\!0$, antichiral superfields by $\!\hat{\,\Da} \bar{\hat\Phi}\!=\!0$
and vector superfields by $\hat V \!=\!\bar{\hat V}$. The component expansion 
is the same as in the classical theory but with noncommutative fields as 
components
\beqa
\hat\Phi (\hat y, \teh) &=& \hat A(\hat y) + \sqrt{2} \teh \hat\psi(\hat y) + 
\teh\teh \hat F(\hat y) \, , \label{eq2.21} \\
\hat V (\hat x,\teh,\tehb) &=& -\teh\sm\tehb \hat \vm(\hat x) + i
\teh\teh\tehb\bar{\hat\lambda} (\hat x) - i \tehb\tehb\teh\hat\lambda (\hat x)
+ \frac{1}{2} \teh\teh\tehb\tehb \hat d(\hat x) \, , \label{eq2.22}
\eeqa 
where $\hat y^{m}=\hat x^{m}+i\teh\sm\tehb$ and $\hat V$ is in the Wess-Zumino 
gauge. 

The supersymmetry transformation of the component fields may be found from 
$\hat\delta_{\hat\xi} \hat\Phi$ and $\hat\delta_{\hat\xi} \hat V$ where
\beq
\hat\delta_{\hat\xi} \equiv \left( \hat\xi \hat Q + \,\bar{\!\hat\xi} \,
\bar{\!\hat Q} \right) \, .
\eeq
Since the generators $\hat Q$ and $\bar{\!\hat Q}$ act on the superfields as 
in the classical case, the transformation of the component fields maintain the 
classical form as well
\beqa
\hat\delta_{\hat\xi} \hat A &=& \sqrt{2} \hat\xi\hat\psi \, , \\ 
\hat\delta_{\hat\xi}\hat\psi &=& i\sqrt{2} \sm\,\bar{\!\hat\xi}\hat\pam \hat A
+ \sqrt{2} \hat\xi \hat F \, , \\
\hat\delta_{\hat\xi}\hat F &=& i\sqrt{2} \,\bar{\!\hat\xi}\sbm\hat\pam\hat\psi
\, ,    \\
\hat\delta_{\hat\xi} \hat\vm &=& i\hat\xi\sdm\bar{\hat\la} - i\hat\la\sdm \,
\bar{\!\hat\xi} \, , \\
\hat\delta_{\hat\xi} \hat\la &=& i\hat\xi\hat d + 2\smn\hat\xi\hat\pam\hat 
v_{n} \, , \\
\hat\delta_{\hat\xi} \hat d &=& \,\bar{\!\hat\xi}\sbm\hat\pam\hat\la - \hat\xi
\sm\hat\pam\bar{\hat\la} \, .
\eeqa

\subsection{Gauge theory}

\noindent
Consider the noncommutative gauge transformation of a chiral superfieldfield 
$\hat\Phi$ 
\beq  \label{eq2.23}
\dlh\hat\Phi = -i\hat\La * \hat\Phi \, , 
\eeq
with the Lie algebra valued noncommutative gauge parameter 
$\hat\La=\hat\La_{a} T^{a}$  and $\hat{\bar D} \hat\La=0$ in order to preserve 
chirality. $T^{a}$ are generators of the appropriate gauge group and form the 
Lie algebra 
\beq \label{eq2.24}
[T^{a},T^{b}] = i f^{ab}_{\,\,\,c} T^{c} \, .
\eeq
The commutator of two gauge transformations has the same form as in the 
classical case
\beq \label{eq2.25}
\dlh \hat\delta_{\hat\Si} - \hat\delta_{\hat\Si} \dlh = 
\hat\delta_{i[\hat\La\stackrel{*}{,}\hat\Si]} \, ,
\eeq
but the commutator 
\beq \label{eq2.26}
[\hat\La\stackrel{*}{,}\hat\Si] = \frac{1}{2} \{\hat\La_{a}\stackrel{*}{,}\hat
\Si_{b}\} [T^{a},T^{b}] + \frac{1}{2} [\hat\La_{a}\stackrel{*}{,}\hat\Si_{b}] 
\{T^{a},T^{b}\} 
\eeq
only closes into the Lie algebra if the gauge group under consideration is 
$U(N)$. Thus in this setting gauge theories with gauge groups $SU(N)$ can not
be considered. However, using Seiberg-Witten map we can consider $SU(N)$ or 
arbitrary groups.  

In what follows we will restrict our considerations to the gauge group $U(1)$.
The generalisation to $U(N)$ is straightforward. The noncommutative 
supersymmetric $U(1)$ gauge theory coupled to matter, which is the 
supersymmetric extension of noncommutative electrodynamics, is constructed in 
terms of two chiral superfields:
\beq
\dlh \hat\Phi_{+} = -i\hat\La * \hat\Phi_{+} \, , \qquad 
\dlh \hat\Phi_{-} = i\hat\Phi_{-} * \hat\La \, .
\eeq
The action is
\beq \label{eq2.27} 
\hat \mathcal{S} = \hat \mathcal{S}_{YM} +  \mathcal{S}_{int_{(+)}} + 
\mathcal{S}_{int_{(-)}} +  \mathcal{S}_{m} \, ,  
\eeq
where
\beqa 
\hat\mathcal{S}_{YM} &=& \frac{1}{4} \int d^{4}x d^{2} \te \, \left( \hat 
W^{\alpha} * \hat W_{\alpha} \right)  + c.c. \, ,  \label{eq2.28} \\
\hat\mathcal{S}_{int_{(+)}} &=& \quad \int d^{4} x d^{2} \te d^{2} \teb\, 
\left( \bar{\hat\Phi}_{+} * e_{*}^{g \hat V} * \hat\Phi_{+} \right) \, ,
\label{eq2.29} \\
\hat\mathcal{S}_{int_{(-)}} &=& \quad \int d^{4} x d^{2} \te d^{2} \teb \,
\left( \hat\Phi_{-} * e_{*}^{-g \hat V} * \bar{\hat\Phi}_{-} \right)  \, ,  \\
\hat\mathcal{S}_{m} &=& m \int d^{4}x d^{2} \te \, \left( \hat\Phi_{-} * 
\hat\Phi_{+} \right) + c.c. \, \label{eq2.30} \, .
\eeqa
The noncommutative exponential function is defined as
\beq \label{eq2.13}
e^{\hat F}_{*} ~=~ \sum_{n=0}^{\infty} \frac{1}{n!} \,(\hat F)^{n}_{*} ~=~ 1 +
\hat F + \frac{1}{2}\, \hat F*\hat F + \frac{1}{6}\, \hat F*\hat F*\hat F + 
\ldots
\eeq

Because of the star product noncommutativity, the gauge transformation of the 
gauge field $\hat V$ has the same form as in the nonabelean case
\beq \label{eq2.31}
\dlh \hat V = -i \bar{\hat\La} * e_{*}^{\hat V} + i e_{*}^{\hat V} * \hat{\La} 
\, .
\eeq
The noncommutative supersymmetric chiral field strength 
\beq \label{eq2.32}
\hat W_{\alpha} = -\frac{1}{4} \bar{\hat D} \bar{\hat D} \,\, e_{*}^{-\hat V} *
\hat D_{\alpha} * e_{*}^{\hat V} 
\eeq
has the gauge transformation
\beq \label{eq2.33}
\dlh \hat W_{\alpha} = i[\hat W\stackrel{*}{,}\hat \La] \, .
\eeq
The expansion of the Lagrangian in component fields in Wess-Zumino gauge has 
the same form as the classical nonabelian supersymmetric gauge theory, e.g
\beq \label{eq2.34}
\mathcal{S}_{YM} = \int d^{4}x \left(-\frac{1}{4} \hat f^{mn} * \hat f_{mn} - i
\bar{\hat\lambda} * \sbm \hat \mathcal{D}_{m} * \hat \lambda +\frac{1}{2} \, 
\hat d*\hat d \right) \, , 
\eeq
where
\beqa
\hat f_{mn} &=& \hat\pam \hat v_{n} - \hat\pan \hat v_{m} + \frac{i}{2} \, [
\hat \vm \stackrel{*}{,} \hat v_{n}] \, , \label{eq2.35} \\
\mathcal{D}_{m} \hat\lambda &=& \hat\pam \hat\lambda + \frac{i}{2} \, [\hat \vm
\stackrel{*}{,} \hat \lambda] \, . \label{eq2.36}
\eeqa

\section{Construction of the Seiberg-Witten map in terms of component fields}

To determine the Seiberg-Witten map in the supersymmetric case we will start 
with the supersymmetric Seiberg-Witten equation for the gauge field $\hat V$ 
and the chiral (matter) field $\hat\Phi$
\beqa \label{eq3.1}
\hat V(V) + \hat\delta_{\hat\Lambda} \hat V(V) &=& \hat V(V + \dl V) \, , 
\label{eq3.1} \\ 
\hat \Phi (\Phi ,V) + \dlh \hat \Phi (\Phi ,V) &=& \hat \Phi (\Psi + \dl\Phi ,
V + \dl V)  \, , \label{eq3.2}
\eeqa
To simplify matters we consider the abelian case, since the generalization to 
nonabelian case is straightforward. We choose furthermore the Wess-Zumino 
gauge in which the equation (\ref{eq2.31}) and the gauge parameter $\hat \La$ 
have the form
\beqa
\dlh \hat V &=& i(\Lah - \Labh) + \frac{i}{2} \, [ \hat V \stackrel{*}{,} \Lah
+ \Labh ] \, . \label{eq3.3} \\
\Lah (y,\te) &=& - \hat\alpha (y) \, , \label{eq3.4} 
\eeqa
where $\hat\alpha$ is the ordinary abelian noncommutative gauge parameter.

For the solution of the Seiberg-Witten equations we use the same procedure as
in the nonsupersymmetric case \cite{Seiberg,Bichl:0102}. We expand the 
superfields in the noncommutative parameter $\tij$
\beqa
\hat V &=& V + V^{'} (V,\vartheta) + o(\vartheta^{2}) \label{eq3.5} \, , \\ 
\hat \Phi &=& \Phi + \Phi^{'}(\Phi,V,\vartheta) + o(\vartheta^{2}) 
\label{eq3.6} 
\, , \\
\Lah &=& \La + \La^{'}(\La,V,\vartheta) + o(\vartheta^{2}) \label{eq3.7} \, ,
\eeqa
and solve the Seiberg-Witten equations (\ref{eq3.1}) and (\ref{eq3.2}) 
perturbatively order by order in the noncommutativity parameter $\tij$. To 
zeroth order we get the classical gauge transformations. To first order we get
\beqa
V^{'} - V^{'}(V + \dl V) + i(\Lap - \Labp) &=& \frac{1}{2} \, \tij \, \pai V 
\paj (\La + \Lab) \, , \label{eq3.8} \\
\Phi^{'} - \Phi^{'} (\Phi + \dl \Phi ,V + \dl V) - i (\Lap \Phi + \La \Phi^{'})
&=& - \frac{1}{2} \, \tij \, \pai \La \paj \Phi \label{eq3.9} \, . 
\eeqa     
These equations can be solved easily componentwise. With the assumption that 
the noncommutative component fields depend only on their classical 
counterparts and the classical gauge field $\vm$, we get the following 
Seiberg-Witten equations for component fields to first order in $\tij$  
\beqa
\vm' - \vm' (v + \dal v ) -2 \, \pam \alp &=&  \tij \pai \al \paj \vm \, , 
\label{eq3.10} \\
\la' - \la' (\la + \dal\la, v + \dal v) &=&  \tij \pai \al \paj \la \, , 
\label{eq3.11} \\
d' - d'(d+\dal d, v +\dal v) &=&  \tij \pai \al \paj d \, , 
\label{eq3.12} \\
A' -A'(A+\dal A,\vm + \dal \vm) +i \al' A +i \al A' &=& \frac{1}{2} \tij 
\pai\al\paj A \, , \label{eq3.13} \\
\psi' -\psi'(\psi+\dal \psi,\vm + \dal \vm) +i \al' A +i \al A' &=& 
\frac{1}{2} \tij \pai\al\paj \psi \, , \label{eq3.14} \\
F' -F'(F+\dal F,\vm + \dal \vm) +i \al' F +i \al F' &=& \frac{1}{2} \tij 
\pai\al\paj F \, , \label{eq3.15} 
\eeqa
The solution is\footnote{We remind that in Wess-Bagger convention, which we 
use, the gauge transformation of the abelian gauge field $\vm$ has the form 
$\dal\vm \!=\!-2\,\pam\al$. This differs from the usual gauge transformation by
the factor $-2$. For this reason the Seiberg-Witten maps for the fields $\al'$,
$\vm'$ and $\psi'$ differ from the usual ones 
\cite{Seiberg,Bichl:0102,Jurco:0104} by the factor $-\frac{1}{2}$.}
\beqa
\al' (\al,v) &=& \frac{1}{4} \tij\vi\paj\al \, , \label{eq3.16} \\
\vm' (v) &=& \frac{1}{4} \tij\vi (\paj\vm + f_{jm}) \, , \label{eq3.17} \\
\la' (\la,v) &=& \frac{1}{2} \tij\vi \paj \la \, , \label{eq3.18} \\
d' (d,v) &=& \frac{1}{2} \tij\vi \paj d \, , \label{eq3.19} \\
A' (A,v) &=& \frac{1}{4} \tij\vi\paj A \, , \label{eq3.20} \\
\psi' (\psi,v) &=& \frac{1}{4} \tij\vi\paj \psi \, , \label{eq3.21} \\
F' (F,v) &=& \frac{1}{4} \tij\vi\paj F \, , \label{eq3.22} 
\eeqa 
The nonabelian generalisation of the solutions (\ref{eq3.16})-(\ref{eq3.19}) 
were also found in \cite{Putz:0205}, where the authors have used a different 
method to determine the Seiberg-Witten equations (\ref{eq3.10})-(\ref{eq3.12}).

Using the above expansions one gets the action (\ref{eq2.27}) expanded to first
order in $\tij$ in terms of classical component fields. For example we get for
the linear term in $\tij$ in the Yang-Mills action (\ref{eq2.34}) 
\beqa 
\mathcal{S}_{YM}^{'} &=& \frac{1}{4} \tij \int d^{4}x \Big\{ -f^{mn} f_{mi} 
f_{nj} + \frac{1}{4}\, f_{ij} f^{mn} f_{mn}   \nonumber  \\ 
&&\qquad\qquad\quad +\, i \big( \pai\la\sm\lab - \la\sm\pai\lab \big) f_{jm}  
\nonumber \\ 
&&\qquad\qquad\quad +\, \frac{i}{2} \left( \pam\la\sm\lab - \la\sm\pam\lab - i
d^{2} \right) f_{ij} \,\, \Big\} \, . \label{eq3.23}
\eeqa
Due to the construction the $\tij$ expanded action is invariant under 
commutative gauge transformations. But, as was realised in \cite{Putz:0205}, it
is not covariant under the commutative supersymmetry transformation. However, 
one can also expand the supersymmetry generators in $\tij$ where the expanded 
terms depend on the commutative component fields of the gauge superfield in 
such a way that the action will be covariant in each order under the action 
of these expanded supersymmetry generators. This was outlined in 
\cite{Putz:0205,Paban:0201} (see also \cite{Dayi:0309}).

The solutions (\ref{eq3.16})-(\ref{eq3.22}) are very restrictive because we 
assumed that the noncommutative component fields depend only on their 
commutative counterparts and the classical gauge field $\vm$. Indeed, there is 
no reason for such a restriction. Giving up this restriction one could ask if 
it is possible to get solutions which will lead to an action which is 
invariant under classical supersymmetry. This question is automatically 
answered by solving equations (\ref{eq3.8}) and (\ref{eq3.9}) explicitly in 
terms of superfields. 

To solve equations (\ref{eq3.8}) and (\ref{eq3.9}) we first have to find  
simultaneously the Seiberg-Witten map for $\La^{'}$ and $V^{'}$. To avoid this
we will apply the method developed by Wess and collaborators in 
\cite{Madore:0012,Jurco:0006,Jurco:0102,Jurco:0104} to determine the 
Seiberg-Witten map for the superfield case.

\section{Construction of the Seiberg-Witten map in terms of Superfields}

We consider  again the noncommutative gauge transformation of a chiral matter 
field (\ref{eq2.23}), but with enveloping algebra valued gauge parameter 
$\hat\La$:
\beq \label{eq4.1}
\hat\La = \La_{a} T^{a} + \Lap_{ab} :T^{a} T^{b}: + \Lapp_{abc} 
:T^{a} T^{b} T^{c}: + \dots \, ,
\eeq
where $\Lap$ is linear in $\tij$, $\Lapp$ is quadratic in $\tij$ etc.
The dots indicate that we have to sum over a basis of the vector space spanned 
by the homogenous polynomials in the generators $T^{a}$ of the Lie algebra. 
Completely symmetrized products could serve as a basis:
\beq \label{eq4.2}
:T^{a}:~=~T^{a} \, , \qquad :T^{a}T^{b}:~=~\frac{1}{2} (T^{a}T^{b}+T^{b}T^{a}) 
\, , \qquad \mathrm{etc}.
\eeq

The commutator of two transformations (\ref{eq2.25}) is certainly enveloping 
algebra valued. Hence we can use arbitrary Lie groups but the price we seem 
to have to pay is an infinite number of gauge parameters and an infinite number
of gauge fields. 

To avoid this problem we define new gauge transformations, where all these  
infinitely many gauge parameters depend just on the classical gauge parameter 
$\La$, the classical gauge field $V$ and on their derivatives. We assume 
moreover that all superfields considered (e.g. $\hat\Phi$, $\hat V$) depend on 
their classical counterpart, the classical gauge field $V$ and on their 
derivatives. This dependence, which we call Seiberg-Witten map, will be 
denoted by $\Lahv$, $\Fihv$ and $\hat V(V)$. 

The gauge transformations for the noncommutative chiral matter field and the 
noncommutative gauge field now have the form 
\beqa 
\dl \Fihv &=& -i \Lahv * \Fihv \, , \label{eq4.3} \\
\dl \hat V(V) &=& -i \bar{\hat\La}(\bar\La,V) * e_{*}^{\hat V(V)} + i 
e_{*}^{\hat V(V)} * \hat{\La}(\La,V) \, . \label{eq4.4}
\eeqa 
Since equation (\ref{eq2.25}), called consistency condition in
\cite{Jurco:0104}, involves solely the gauge parameters, it is convenient to 
base the construction of the Seiberg-Witten map on it. In a second step the 
remaining Seiberg-Witten maps for the matter field and the gauge field can be
computed from the equations (\ref{eq4.3}) and (\ref{eq4.4}).
 
The procedure in the abelian case is the following. As was mentioned, we start
with the consistency condition which has the following form in the abelian case
\beq \label{eq4.5}
(\dl \ds - \ds \dl) \, \Fihv = 0 \, .
\eeq  
With equation (\ref{eq4.3}) we get more explicitly
\beq \label{eq4.6}
-i \dl \Sihv + i \ds \Lahv - \Sihv * \Lahv + \Lahv * \Sihv = 0 \, .
\eeq
The variation $\dl \Sihv$ refers to the V-dependence of $\Sihv$ and the gauge 
transformation of the supersymmetric abelian gauge field V 
\beq \label{eq4.7}
\dl V = i (\La - \Lab) \, .
\eeq
We now expand the consistency condition and the gauge transformations 
(\ref{eq4.3}) and (\ref{eq4.4}) using the expansions 
(\ref{eq3.5})-(\ref{eq3.7}) and the expanded star product (\ref{eq2.9}). From
these expanded equations we can then determine the Seiberg-Witten maps order by
order in $\tij$ for all considered superfields. To first order in $\tij$ we 
get the following equations 
\beqa
\dl \Sipv - \ds \Lapv &=& \tij \pai \La \paj \Si \, , \label{eq4.8}\\
\dl \Fipv +i \La\Fipv +i \Lapv\Phi &=& \frac{1}{2} \tij \pai \La \paj \Phi 
\, , \label{eq4.9}\\
\dl V^{'} (V) - i\Lapv +i\Labpv &=& \frac{1}{2} \tij\pai (\La + \Lab) \paj V 
\, . \label{eq4.10}
\eeqa
We will now look for solutions of these equations.

\section{Seiberg-Witten map on Minkowski space}

It is reasonable to require the following properties of the Seiberg-Witten map 
for superfields:
\begin{enumerate}
\item
locality (in each order),
\item 
chirality of the gauge parameter \\
(otherwise we would not have a gauge theory which involves only one real 
vector superfield $\hat V$ since it would not be possible to impose the so 
called \textit{representation preserving constraints} 
\cite{Gates:1979,Gates:1980}),
\item
covariance under classical supersymmetry.
\end{enumerate}
First we will show that on canonically deformed Minkowski superspace there is 
no Seiberg-Witten map for $\La^{'}$ which fulfils all three requirements. 
However, as we will see, there are solutions if we give up one of the 
requirements.

\subsection{No go theorem} \label{sec:nogo}

\textbf{Theorem:} \textit{On canonically deformed Minkowski superspace 
(see equation (\ref{eq2.1})) there is no Seiberg-Witten map for the gauge 
parameter $\hat\La$ which is local, chiral and supersymmetric at the same 
time.} \\

To prove this theorem, it is sufficient to prove that there is no local, 
chiral and supersymmetric solution of the consistency condition to first order 
in $\tij$ (\ref{eq4.8}). We show this by using dimensional analysis.

The right hand side of the consistency condition (\ref{eq4.8}) is linear in 
each of the classical superfields $\La$ and $\Si$. All terms in the ansatz for 
$\La^{'}$ which would contain powers of $V$ can therefore solve only the 
homogeneous consistency condition because of (\ref{eq4.7}). Hence we make an 
ansatz for $\La^{'}$ only linear in $V$ without loss of generality. Moreover 
$\La^{'}$ has to be linear in the classical gauge parameter $\La$ and by 
definition linear in $\tij$. In order to have a supersymmetric expression we 
may use the bosonic derivatives $\pam$ and the covariant derivatives $\Da$ and 
$\Dba$ only. The mass dimensions of these objects are
\beq \label{eq5.1}
[\La^{'}] = [\La] = [V] = 0 \, , \quad [\tij] = -2 \, , \quad [\pam] = 1 \, ,
\quad [\Da] = [\Dba] = \frac{1}{2} \, .  
\eeq \label{eq5.2}
It is not hard to see that there is only one local, chiral and supersymmetric 
combination of $\tij$, $\La$, $V$, $\partial_{m}$, $\Da$ and $\Dba$ 
with appropriate index contraction and mass dimension zero. It is 
\cite{Ferrara}
\beq \label{eq5.2}
\La^{'} = a P^{\alpha\beta} \bar D^{2} (\Da\La\Db V) \, ,
\eeq
where $a$ is a constant and 
\beq \label{eq5.3}
P^{\alpha\beta}=\vartheta_{ij} \varepsilon^{\alpha\mu} 
(\sigma^{ij})_{\mu}^{\,\,\,\beta} \, .  
\eeq
We put this ansatz for $\La^{'}$ and $\Si^{'}$ in the left hand side of the 
consistency condition (\ref{eq4.8}) and get
\beqa 
\dl \Si^{'} - \ds \La^{'} &=& 32ai Tr\left(\sigma^{ij}\sigma^{mn}\right) 
\pam\La\pan\Si \nonumber \\ 
&=& - 32ai \,\left( \tij + \frac{i}{2} 
\varepsilon^{ijmn} \vartheta_{mn}\right) \pai\La\paj\Si \, .\label{eq5.4}
\eeqa
There is no choice of the constant $a$ which makes this expression equal to 
the right hand side of the consistency condition. Thus we proved the above 
theorem. 

Nevertheless, it is obvious from equation (\ref{eq5.4}) that the ansatz
(\ref{eq5.2}) would be a solution of the consistency condition if we would 
take $\tij$ to be complex and selfdual. But in this case we would have to 
complexify the canonically deformed Minkowski superspace. The other possibility
is to consider the canonically deformed Euclidean superspace on which the 
selfdual $\tij$ is real. Before we do this, let us consider the 
Seiberg-Witten maps for superfields on deformed Minkowski superspace when one 
of the above requirements is given up.

\subsection{Nonchiral solution} \label{sec:nonchiral}

A nonchiral solution of the consistency condition (\ref{eq4.8}) is 
\beq \label{eq5.5}
\Lapv = - \frac{1}{16} \Paabb (\daad \La) (\Dbb\Db V) \, ,
\eeq 
where $\Paabb$ and $\daad$ are the noncommutative parameter and the bosonic 
derivative in spinor notation
\beq \label{eq5.6}
\Paabb = \vartheta_{ij} \bar\sigma^{i\,\dot\alpha\alpha} \bar
\sigma^{j\,\dot\beta\beta} \, , \qquad \daad = \sm_{\alpha\dot\alpha} \pam \, .
\eeq
One gets this nonchiral Seiberg-Witten map for the gauge parameter immediately
by the attempt to write the equation 
\beq \label{eq5.7}
\Lap \big|_{\te=\teb=0} = -\alp = -\frac{1}{4} \tij\vi\paj\al 
\eeq
in a supersymmetric way in terms of the superfields $\La$ and $V$ since
\beqa 
\daad \La \big|_{\te=\teb=0} &=& -\sm_{\alpha\dot\alpha} \pam\al\, , 
\label{eq5.8} \\
\Dbb\Db V \big|_{\te=\teb=0} &=& \sm_{\beta\dot\beta} \,\vm \, . \label{eq5.9}
\eeqa
With this solution we can now solve the equations (\ref{eq4.9}) and 
(\ref{eq4.10}) and obtain the following Seiberg-Witten maps for $\Phi^{'}$ and 
$V^{'}$:
\beqa
\Fipv &=& -\frac{1}{16} \Paabb (\daad \Phi) (\Dbb\Db V) \, , \label{eq5.10} \\
V^{'}(V) &=& -\frac{1}{16} \Paabb (\daad V) (\Dbb\Db V) \, . \label{eq5.11} 
\eeqa 
The solution (\ref{eq5.10}) is nonchiral as well. 

In the Wess-Zumino gauge we get the following component expansion of these
solutions:  
\beqa
\Lap(y,\te,\teb) &=& -\frac{1}{4} \tij \bigg\{ 
\vi\paj\al + i\te\si\lab\paj\al + i\teb\sbi\la\paj\al  \nonumber\\
&&\quad\qquad -i\te\sm\teb \left[ \Big( f_{jm}^{SD} -i \eta_{mj} d
\Big) \pai \al \right]  \nonumber \\ 
&&\quad\qquad +\te\te\teb  \sbj\sm\pam\lab\pai\al    \bigg\} \, ,
\label{eq5.12}
\eeqa
\beqa
\Phi(y,\te,\teb) &=& - \frac{1}{4} \tij \bigg\{ \pai A \vj + \sqrt{2}\te \Big( 
\pai\psi\vj + \frac{i}{\sqrt{2}} \sj\lab\pai A \Big) \nonumber \\ 
&&\quad\qquad + i\teb\sbj\la\pai A + \te\te\left[ \pai F \vj - 
\frac{i}{\sqrt{2}} \pai\psi\sj\lab \right]  \nonumber \\
&&\quad\qquad + i\te\sm\teb\bigg[ \Big( f_{jm}^{SD} -i \eta_{mj} d
\Big) \pai A - \frac{i}{\sqrt{2}} \la\sj\sbdm\pai\psi \bigg]  \nonumber \\    
&&\quad\qquad + \te\te\teb \bigg[ \frac{i}{\sqrt{2}} \Big( \sbm f_{jm}^{SD}  -
 i \sbj d \Big) \pai \psi  \nonumber \\  
&&\quad\qquad \qquad\quad + i\sbj\la\pai F - \sbj\sm\pam\lab\pai A \bigg]  
\bigg\} \, , \label{eq5.13}
\eeqa
\beqa
V^{'} (x,\te,\teb) &=& \frac{1}{4} \tij \bigg\{ \te\sm\teb \left( \vj\pai\vm 
\right) - i\te\te\teb \left[ \pai\lab\vj - \frac{1}{2} \sbm\sj\lab\pai\vm 
\right]  \nonumber \\
&&\quad\qquad + i\teb\teb\te \left[ \pai\la\vj - \frac{1}{2} \sm\sbj\la\pai\vm 
\right]  \nonumber \\
&&\quad\qquad - \frac{1}{2} \te\te\teb\teb \left[ \pai(\vj d) - \pai v^{m}\,
{}^{\star}\!f_{jm}  - \pai(\la\sj\lab) \right] \bigg\} \, , 
\label{eq5.14}
\eeqa
where ${}^{\star}\!f_{ab}=\frac{1}{2}\varepsilon_{abcd}f^{cd}$ and 
$f_{mn}^{SD}$ is the complexifyed selfdual field strength defined as usual:
$f_{mn}^{SD}=f_{mn}+i\,{}^{\star}\! f_{mn}$. 

From the component expansion we see that there are additional peculiar aspects
of this solution besides the nochirality of the gauge parameter. From
(\ref{eq5.14}) we read off the Seiberg-Witten map for the ordinary gauge field
$\vm'$
\beq  \label{eq5.15}  
\vm' = \frac{1}{4} \tij \vi\paj\vm \, ,
\eeq
and realise that it does not coincide with the original one of Seiberg and
Witten which is (\ref{eq3.17}). Furthermore, from the equation (\ref{eq5.12}) 
we see that the $-i\te\sm\teb$ component of $\La^{'}(x,\te,\teb)$ is 
\beq \label{eq5.16}
\La^{'}(x,\te,\teb) \big|_{-i\te\sm\teb} = -\pam\al' + f_{m}' ,   
\eeq
where
\beqa
\al'(\al,v) &=& \frac{1}{4} \tij\vi\paj\al \, , \label{eq5.17}  \\
f_{m}'(\al,v,d) &=& -\frac{1}{4}\tij \left( f_{jm}^{SD} -i \eta_{mj} d \right)
 \pai \al \, . \label{eq5.18}
\eeqa
This implements an unusual gauge transformation of the noncommutative abelian 
gauge field $\hat\vm$  
\beq \label{eq5.19}
\dal \hat\vm = - 2 \hat\pam\hat\al + i[\hat\al\stackrel{*}{,}\hat\vm] -  
(\hat f_{m} + \!\!\bar{{\,\,\hat f}_{m}}) \, .
\eeq

We want to make one more comment on the nonchirality of the superfield 
$\hat\La$. Since $\La^{'}$ is a complex linear superfield satisfying the
constraint $\bar D^{2}\La^{'}=0$, one could ask if the same is true for 
$\hat \La$. This would be the case if the Seiberg-Witten map to all orders 
would satisfy this constraint. With some computation one can show that
\beq \label{eq5.21}
\La{''} (\La,V) = - \frac{i}{2^{8}} \Paabb P^{\dot\mu\mu\dot\nu\nu} 
(\Dba\Da\La) (\bar D_{\dot\mu} D_{\mu} V) \partial_{\beta\dot\beta} 
(\bar D_{\dot\nu} D_{\nu} V)
\eeq
solves the consistency condition expanded to second order in the 
noncommutative parameter
\beq \label{eq5.20}
\dl \Si^{''} -\ds \La^{''} = \tij\pai\La\paj\Si^{'} -\tij\pai\Si\paj\La^{'}\, .
\eeq 
This solution does not satisfy the constraint $\bar D^{2}\La^{''}=0$ and is 
therefore not a complex linear superfield. Since this solution is not unique,
it is still an open question if there exists a complex linear solution to 
second order.

\subsection{Nonlocal solution}

The most natural way to create a chiral Seiberg-Witten map for the 
supersymmetric gauge parameter out of the nonchiral one (\ref{eq5.5}) is to 
insert the chiral projector 
\beq \label{eq5.22}
P = \frac{\bar D^{2}D^{2}}{16\Box} \, ,
\eeq
where $\Box\!=\!\partial^{m}\pam$ in front of the nonchiral term $\Dba\Da V$. 
The Seiberg-Witten map created in this way
\beq \label{eq5.23}
\Lapv = - \frac{1}{16} \Paabb (\daad \La) P (\Dbb\Db V) \, ,
\eeq  
is still a solution of the consistency condition (\ref{eq4.8}). 

The nonlocality enters in this solution due to the nonlocal chiral projector. 
As a matter of fact, the chiral constraint $\Dba\Lap\!=\!0$ contains 
derivatives and is like a differential equation in superspace. Hence, it is not
ultralocal in the $x$ variables (ultralocality means that the constraint is
determined exclusively by the function at $x$, but not it's derivatives).
This is why the chiral projectors on fields are nonlocal. One has to invert 
derivatives with respect to $x$.

The corresponding nonlocal solutions of the equations (\ref{eq4.9}) and 
(\ref{eq4.10}) are 
\beqa
\Fipv &=& -\frac{1}{16} \Paabb (\daad \Phi) P (\Dbb\Db V) \, . \label{eq5.24}\\
V^{'}(V) &=& -\frac{1}{16} \Paabb (\daad V) P (\Dbb\Db V) + \frac{1}{16} 
\Paabb (\daad V) \bar P (\Db\Dbb V)  \nonumber \\
&& - \frac{i}{32} \Paabb (P\Dba\Da V)(\bar P \Db\Dbb V) \, . \label{eq5.25}    
\eeqa 
It is obvious that $\Dba\Phi^{'}=0$. Since $P\Dba\Da = -2iP\partial_{\alpha\dot
\alpha}$ we can write the Seiberg-Witten maps (\ref{eq5.23})-(\ref{eq5.25}) in 
a simpler form
\beqa
\Lapv &=& \frac{1}{2} \tij \pai\La P\paj V \, , \label{eq5.26} \\
\Fipv &=& \frac{1}{2} \tij \pai\Phi P\paj V \, , \label{eq5.27} \\
V^{'}(V) &=& \frac{i}{2} \tij (1-P)\pai V (1-\bar P) \paj V \, . \label{eq5.28}
\eeqa
The solutions (\ref{eq5.26}) and (\ref{eq5.28}) were first found in 
\cite{Chekhov:2001}. 

The nonlocal Seiberg-Witten maps for the component fields can be deduced from 
the nonsupersymmetric Seiberg-Witten maps that we will now discuss.

\subsection{Nonsupersymmetrtic solution}

Another possibility in order to modify equation (\ref{eq5.5}) into a chiral one
is to use the nonsupersymmetric chiral projector $M$
\beq \label{eq5.29}
M = \frac{1}{16\Box} \bar D^{2} D^{\alpha} M_{\alpha} \, ,
\eeq
where
\beq
M_{\alpha} = -\partial_{\alpha} + 3i \left( \sm\teb \right)_{\alpha} \pam \, ,
\eeq
instead of the supersymmetric one (\ref{eq5.22}). The nonsupersymmetric  
Seiberg-Witten maps for the fields $\La^{'}$, $\Phi^{'}$ and $V^{'}$ are 
therefore the same as (\ref{eq5.23})-(\ref{eq5.25}), where the chiral 
projector $P$ is replaced by $M$ and the antichiral projector $\bar P$ by 
$\bar M$, respectively.

It is worth noticing that $Q_{\alpha}$ commutes with $M$ and 
$\bar Q_{\dot\alpha}$ with $\bar M$. The chiral fields $\Lap$ and $\Phi^{'}$ 
retain therefore $N\!=\!(\frac{1}{2},0)$ and the antichiral fields $\bar\Lap$ 
and $\bar\Phi^{'}$
$N\!=\!(0,\frac{1}{2})$ of the original $N\!=\!(\frac{1}{2},\frac{1}{2})$ 
supersymmetry.\footnote{Recently, Seiberg \cite{Seiberg:0305} considered the
$\te-\te$ deformed superspace which keeps the $N\!=\!\frac{1}{2}$ 
supersymmetry 
and has some interesting properties from the field theoretical viewpoint (see 
e.g. \cite{Berenstein:0308} and references therein). Therefore it is more 
suitable to denote the $N\!=\!1$ superspace as 
$N\!=\!(\frac{1}{2},\frac{1}{2})$ superspace.} 
The vector superfield $V^{'}$ however brakes the entire 
$N\!=\!(\frac{1}{2},\frac{1}{2})$ supersymmetry because it contains both 
projectors.
   
The fields $\La^{'}$ and $V^{'}$ have the following component expansion in the 
Wess-Zumino gauge 
\beqa
\La^{'} (y,\te) &=& - \, \al' (y) + \te\eta'(y) \, , \label{eq5.30} \\
V^{'} (x,\te,\teb) &=& i \te\chi'(x) -  i \teb\bar\chi'(x) - \, \te\sm\teb 
\vm' (x) + i \te\te\teb \left[ \lab' (x) + \frac{i}{2} \sbm
\pam \chi' (x) \right]  \nonumber \\   
&& - i \teb\teb\te \left[ \la' (x) + \frac{i}{2} \sm \pam\bar\chi'(x) \right] 
+  \frac{1}{2}  \te\te\teb\teb d'(x) \, . \label{eq5.31}
\eeqa
These are different from the classical ones (\ref{eq3.4}) and (\ref{eq2.22}), 
while the component expansion of $\Phi^{'}$ is still the classical one 
(\ref{eq2.21}). The Seiberg-Witten maps for the component fields are
\beqa 
\al' (\al,v) &=& - \frac{1}{4} \, \tij \pai\al\vj\, ,\label{eq5.32} \\
\eta' (\al,\lab) &=& - \frac{i}{4} \, \tij \si \bar \la \paj \al\, ,
\label{eq5.33} \\ \nonumber \\
A^{'} (A,v) &=&  - \frac{1}{4} \, \tij  \pai A \vj  \, , \label{eq5.34}\\
\psi{'} (\psi,v,\lab) &=& - \frac{1}{4} \, \tij \left( \pai 
\psi \vj + \frac{i}{\qw} \, \sj \lab \pai A \right) \, ,
\label{eq5.35}  \\
F^{'} (F,\psi,v,\lab) &=& - \frac{1}{4} \tij \, \left( \pai F \vj - 
\frac{i}{\qw} \, \pai\psi\sj\lab \right) \, , \label{eq5.36} \\ \nonumber  \\
\chi' (\lab,v) &=& \,\,\,\, \frac{i}{8} \, \tij \si\lab\vj\, ,\label{eq5.37}\\ 
\vm^{'} (v,\la,\lab) &=& \,\,\,\, \frac{1}{4} \, \tij \left( \vi \left( 
\paj\vm + f_{jm}\right) + \frac{1}{4} \, \varepsilon_{ijmn} \la\sn\lab \right)
 \, ,  \label{eq5.38} \\
\la^{'} (\la,v) &=& - \frac{1}{4} \, \tij \left( \pai\la\vj - \frac{1}{2} \, 
\sm\sbi\la f_{mj}  \right)   \, , \label{eq5.39} \\
d^{'} (d,v,\la,\lab) &=& - \frac{1}{4} \, \tij \,\bigg( 2\pai d\vj - \, 
\pai\la\sj\lab + \frac{i}{2}\, {}^{\star}\!\!\left(\pai\la\sj\lab\right)  
\nonumber \, \\
&& \qquad \qquad - \la\sj\pai\lab  - \frac{i}{2}\, {}^{\star}\!\!\left( 
\la\sj\pai\lab \right) \bigg) \, , \label{eq5.40}
\eeqa
where ${}^{\star}\!\!\left(\pai\la\sj\lab\right)=\frac{1}{2}\varepsilon_{ijmn}
\paum\la\sigma^{n}\lab$.

Comparing these solutions with the solutions obtained in chapter 3, we see 
that additional terms enter which are solutions of the homogenous 
Seiberg-Witten equations. Hence, these solutions are connected to the solutions
in chapter 3 via field redefinition. Furthermore, the Seiberg-Witten maps 
(\ref{eq5.32}), (\ref{eq5.35}) and (\ref{eq5.38}) coincide with the original
ones proposed by Seiberg and Witten when the superpartner fields are set to 
zero.

The expansion of the field strength in classical component fields can be 
determined from equations (\ref{eq2.32}), (\ref{eq5.31}) and  
(\ref{eq5.37})-(\ref{eq5.40}). It is
\beq \label{eq6.4}
W^{'}_{\alpha} (y,\te,\teb) = \frac{1}{4} \, \tij W_{\alpha\,ij} \, ,
\eeq
where
\beqa 
W_{\alpha\,ij}  &=&  
2i \, \pai\la_{\alpha} \vj + \frac{i}{2} \, \big( \sm\sbi\la \big)_{\alpha}
 f_{jm}  \nonumber  \\ 
&-& \!\!\!\! \te_{\alpha} \left[ 2\pai d\vj - \, \pai\la\sj\lab + \frac{i}{2}
\, {}^{\star}\!\!\left(\pai\la\sj\lab\right) - \la\sj\pai\lab  - \frac{i}{2}\,
{}^{\star}\!\!\left( \la\sj\pai\lab \right) \right]  \nonumber \\ 
&+& \!\!\!\!  2i\, \big( \smn\te \big)_{\alpha} \left[ f_{mi} f_{nj} - 
\vi\paj f_{mn} - \frac{i}{8} \,\pam (\la\si\sbdn\sj\lab) + \frac{i}{8} \,
\pan \left( \la\si\sbdm\sj\lab \right) \right]  \nonumber \\
&+& \!\!\!\!  \te\te \left[ 2 \left( \sm\pai\lab \right)_{\alpha} f_{jm} - 
2 \left( \sm\pam\pai\lab \right)_{\alpha} \vj + \frac{1}{2} \left( 
\sm\sbn\si\pam \left( \lab f_{nj} \right) \right)_{\alpha}  \right] \! . 
\label{eq6.5}  
\eeqa

Now we want to make a remark on the Seiberg-Witten map for the component fields
of the nonlocal solutions. The difference lies in using different projectors.
The difference in the component expansion comes from the terms on which the 
projectors act. These terms are
\beqa
\left( M \bar D \sj D V \right) (y,\te) &=& -2 \vj - 2i \te\sj\lab \, , 
\label{eq5.41} \\
\left( P \bar D \sj D V \right) (y,\te) &=& -2 \left( \frac{\paj\pam}{\Box} 
v^{m} - i \frac{\paj}{\Box} d \right) + 4i \te\sm \frac{\paj\pam}{\Box} \lab 
\, . \label{eq5.42}   
\eeqa
From this we see that we get the Seiberg-Witten map for the component fields of
the nonlocal solution by the following  replacement of each $\vj$ and 
$\sj\lab$ (respectively $\vi$ and $\si\lab$) in equations (\ref{eq5.32}) - 
(\ref{eq5.40}):
\beqa
\vj &\longrightarrow& \frac{\paj\pam}{\Box} v^{m} - i \frac{\paj}{\Box} d 
\, , \label{eq5.43}\\
\sj\lab &\longrightarrow& -2 \sm \frac{\paj\pam}{\Box} \lab \, . \label{eq5.44}
\eeqa

\subsection{Nonsupersymmetric noncommutative electrodynamics}

We are now able using the Seiberg-Witten maps (\ref{eq5.34}-\ref{eq5.40}) to 
expand the action (\ref{eq2.27}) in classical component fields to first order 
in $\tij$. With the expansion of the superfields and star product we get
\beqa 
\mathcal{S}_{YM}^{'} &=& \frac{1}{2} \int d^{4}x d^{2} \te \,  \left(
W^{'\alpha} W_{\alpha} \right) + c.c. \, , \label{eq6.1} \\ 
\mathcal{S}_{int_{(+)}}^{'} &=& \quad \int d^{4} x d^{2} \te d^{2} \teb \, 
\Bigg\{ \Fik\Fis + 
e\left( \Fik V\Fis + \frac{i}{4} \, \tij\Fik\pai V\paj\Phi_{+} \right) \, 
\nonumber \\ 
&& \!\!\!\!\! + \, \frac{e^{2}}{2} \left( \Fik V^{2}\Fis + \Fik V V^{'}\Phi_{+}
+ \frac{i}{4} \, \tij\Fik\pai V^{2}\paj\Phi_{+} \right) \Bigg\} + c.c. \, , 
\label{eq6.2} \\ 
\mathcal{S}_{m}^{'} &=& m \int d^{4}x d^{2} \te \, \left( \Phi_{-}^{'}\Phi_{+}
+ \Phi_{-}\Phi_{+}^{'} \right) + c.c. \, . \label{eq6.3}
\eeqa
We omit the action $\mathcal{S}_{int_{(-)}}^{'}$ since it can be easily deduced
from the action $\mathcal{S}_{int_{(+)}}^{'}$. The Seiberg-Witten maps for the
component fields of the chiral superfields $\Phi_{+}^{'}$ and $\Phi_{-}^{'}$ 
have the same form (\ref{eq5.34})-(\ref{eq5.36}).  The action (\ref{eq2.27})
expanded in classical components via Seiberg-Witten map to first order in 
$\tij$ is
\beqa 
\mathcal{S}_{YM}^{'} &=& \mathrm{Equation\,\, (\ref{eq3.23})}  \nonumber \\ 
&+& \frac{1}{4} \int d^{4}x\, \tij \bigg\{ \Big[\pai\la\sj\lab - \frac{i}{2}
\, {}^{\star}\!\!\left(\pai\la\sj\lab\right) + \la\sj\pai\lab  + \frac{i}{2}\,
{}^{\star}\!\!\left( \la\sj\pai\lab \right) \Big] d  \nonumber \\  
&&\qquad \qquad  +\, \frac{i}{2} \Big[ \la\si\sbm\sn \partial_{n} \lab -
 \partial_{n} \la \sn \sbm \si \lab \Big] f_{jm} \bigg\} \nonumber \\ 
&=& \frac{1}{4} \int d^{4}x\, \tij \bigg\{ -f^{mn} f_{mi} 
f_{nj} + \frac{1}{4}\, f_{ij} f^{mn} f_{mn} + \frac{1}{2}\, d^{2} f_{ij} 
\nonumber  \\ 
&&\qquad \quad  +\, \frac{i}{2} \Big[\pai\la\sm\lab + i
\, {}^{\star}\!\!\left(\pai\la\sm\lab\right) - \la\sm\pai\lab  + i \,
{}^{\star}\!\!\left( \la\sm\pai\lab \right)   \nonumber \\ 
&&\qquad \quad  +\paum\la\si\lab - i
\, {}^{\star}\!\!\left(\paum\la\si\lab\right) - \la\si\paum\lab  - i \,
{}^{\star}\!\!\left( \la\si\pam\lab \right) \Big] f_{jm} \nonumber \\ 
&&\qquad \quad + \Big[\pai\la\sj\lab - \frac{i}{2}
\, {}^{\star}\!\!\left(\pai\la\sj\lab\right) + \la\sj\pai\lab  + \frac{i}{2}\,
{}^{\star}\!\!\left( \la\sj\pai\lab \right) \Big] d \bigg\} \, , \label{eq6.6}
\eeqa
\beqa 
\mathcal{S}_{int_{(+)}}^{'} &=& \frac{1}{4} \int d^{4} x\, \tij \bigg\{ 
\paum\vfs\pam (\pai\vf v_{j}) + \paum\vf\pam 
(\pai\vfs v_{j}) - \frac{1}{2} F\bar F f_{ij}  \nonumber \\ 
&& \!\!\!\!\! +\, i \big( \pai\psi\sm\pam\psib - \pam\psi\sm\pai\psib \big) 
v_{j} + \frac{i}{\sqrt{2}} \big( \bar F\pai\psi\sj\lab - F \la\sj\pai\psib 
\big)  \nonumber \\ 
&& \!\!\!\!\! +\, \frac{1}{\sqrt{2}} \big( \pam\psib\sbm\sj\lab\pai\vf +
\la\sj\sbm\pam\psi\pai\vfs \big)  \nonumber \\
&& \!\!\!\!\! +\,\frac{e}{2} \, \Big[ \quad \,\, 2\, \big( \paj\vfs\paum\vf +
\paum\vfs\paj\vf \big) \pai\vm - 2i \vfs\paj\vf\pai d  \nonumber \\ 
&& \qquad +\, 2\sqrt{2} \big( \vfs\pai\la\paj\psi + \vf\pai\lab\paj\psib 
\big) +2i \pai\vm\paj\psi\sm\psib  \nonumber \\ 
&& \qquad +\, i \big( \vfs\pai\vf - \vf\pai\vfs \big) \vj\paum\vm - \big( 
\vfs\pai\vf + \vf\pai\vfs \big) \big( d\vj - \la\sj\lab \big)   \nonumber \\
&& \qquad +\, 2i \big( \paum\vfs\pai\vf - \paum\vf\pai\vfs - \frac{i}{4} \,
\pai\psi\sm\psib - \frac{i}{4} \, \psi\sm\pai\psib \big) \vj\vm  \nonumber \\
&& \qquad +\, i\sqrt{2} \big( \vf\lab\pai\psib - \vfs\la\pai\psi + 
\pai\vf\lab\psib - \pai\vfs\la\psi \big) \vj  \nonumber \\
&& \qquad +\, \frac{i}{\sqrt{2}} \big( \pai\vfs\la\sj\sbm\psi - 
\pai\vf\psib\sbm\sj\lab \big) \vm \quad \,\, \Big]  \nonumber \\
&& \!\!\!\!\! +\,\frac{e^{2}}{4} \Big[ \quad \,\, 2i \pai\vfs\paj\vf\vum\vm + 
\big( \vfs\pai\vf + \vf\pai\vfs \big) \vj\vum\vm  \nonumber \\
&& \qquad -\, 2\vfs\vf\vum \big[ \vi (\paj\vm + f_{jm}) + \frac{1}{4} \,
\varepsilon_{ijmn} \la\sn\lab \big]  \nonumber \\
&& \qquad +\, \frac{1}{\sqrt{2}} \big( \vfs\psi\sm\sbj\la + \vf\lab\sbj\sm\psib
\big) \vi\vm \quad \,\, \Big] \bigg\}\, , \label{eq6.7}
\eeqa

\beqa 
\mathcal{S}_{m}^{'} &=& -\frac{m}{8} \int d^{4}x\, \tij \bigg\{ \Big( \psi_{+}
\psi_{-} -  A_{+}F_{-} - A_{-}F_{+} \Big) f_{ij}  \nonumber \\ 
&& \qquad \quad + i\qw \Big( A_{+} \psi_{-} + A_{-} \psi_{+} \Big)
\sj\pai\lab \, \bigg\} \, + \, c.c.
\eeqa
Since all component fields in $\mathcal{S}_{int_{(+)}}^{'}$ have the same lower
index we have omitted it to make the formula more readable.

\section{Seiberg-Witten map on Euclidean space}

We will be working on canonically deformed Euclidean superspace, but we will
continue to use Lorentzian signature notation. We pass from Minkowski to 
Euclidean space in the convenient way by substituting $x^{0}\to -ix^{0}$. 
Although the formulas in section 2 still hold we should bear in mind that 
dotted and undotted spinors, e.g. $\te$ and $\teb$, are independent objects on 
Euclidean space. This applies to chiral and antichiral fields, too. 
  
The defining equations for the Seiberg-Witten map remain the same 
(\ref{eq4.8})-(\ref{eq4.10}). Thus the nonchiral, nonlocal and 
nonsupersymmetric Seiberg-Witten maps obtained in the previous section on 
Minkowski space remain unchanged on Euclidean space. 

Let us now take the noncommutativity parameter to be selfdual
${}^{*}\tij\!=\!\tij$ and denote it by $\tij_{SD}$ in what follows. In this 
case there exists a local, chiral and supersymmetric solution of the 
consistency condition (\ref{eq4.8}) with $\tij$ replaced by $\tij_{SD}$. It is
\beq \label{eq7.1}
\La^{'} = \frac{i}{64} P^{\alpha\beta}_{SD} \bar D^{2} (\Da\La\Db V) \, ,
\eeq 
where\footnote{It doesn't matter if we use a general $\tij$ or 
$\vartheta_{ij}^{SD}$ in (\ref{eq7.2}) since $\vartheta_{ij}^{SD}\sigma^{ij}=2
\vartheta_{ij}\sigma^{ij}$ due to the selfduality of $\sigma^{ij}$.}    
\beq \label{eq7.2}
P^{\alpha\beta}_{SD} = \vartheta_{ij}^{SD} \varepsilon^{\alpha\mu} 
(\sigma^{ij})_{\mu}^{\,\,\,\beta} \, .
\eeq
With some calculation one can show that the solution (\ref{eq7.1}) can be 
written as
\beq \label{eq7.3}
\La^{'} = \La^{'}_{nonchiral} + \La^{'}_{hom} \, , 
\eeq 
where $\La^{'}_{nonchiral}$ is the nonchiral solution (\ref{eq5.5}) with 
$\Paabb$ replaced by
\beq \label{eq7.4}
\Paabb_{SD} = \vartheta_{ij}^{SD} \bar\sigma^{i\,\dot\alpha\alpha} \bar
\sigma^{j\,\dot\beta\beta} \, .
\eeq
The second term 
\beq \label{eq7.5}
\La^{'}_{hom} = -\frac{i}{16} P^{\alpha\beta}_{SD} \Da\La W_{\beta}
\eeq
is a solution of the homogenous consistency condition.

The corresponding Seiberg-Witten map for the field $\Phi^{'}$ is
\beq \label{eq7.6}
\Fipv = -\frac{1}{16} \Paabb_{SD} (\daad \Phi) (\Dbb\Db V) -\frac{i}{16} 
P^{\alpha\beta}_{SD} \Da\La W_{\beta} \, . 
\eeq
In order to determine the Seiberg-Witten map for the gauge superfield $\hat V$,
we have first to ascertain the Seiberg-Witten map for the superfield 
$\bar{\hat\La}$. On the Minkowski space this can be easily done by taking the 
complex conjugate of $\hat\La$, whereas on Euclidian space in we have to 
solve the consistency condition for $\bar{\hat\La}$ explicitly. In the abelian 
case the consistency condition for $\bar{\hat\La}$ to first order in 
$\vartheta_{ij}^{SD}$ is
\beq \label{eq7.7}
\dl \Sibpv - \ds \Labpv = \vartheta_{ij}^{SD} \pai \bar\La \paj \bar\Si \, .
\eeq

From section \ref{sec:nogo} it is obvious that the only local, antichiral and 
supersymmetric ansatz for the Seiberg-Witten map for $\bar\La^{'}$ is
\beq \label{eq7.8}
\bar\La^{'} =a \bar P^{\dot\alpha\dot\beta}_{ASD} D^{2} (\Dba\La\Dbb V) \, ,
\eeq   
with
\beq \label{eq7.9}
\bar P^{\dot\alpha\dot\beta}_{ASD} = \vartheta_{ij}^{ASD} 
\varepsilon_{\dot\beta\dot\mu} (\bar\sigma^{ij})^{\dot\alpha}_{\,\,\,\dot\mu} 
\, .
\eeq
This is however not a solution of the consistency condition (\ref{eq7.7}). But 
it is obvious that it would be a solution if we take the noncommutativity 
parameter to be antiselfdual. Thus if the deformation parameter is selfdual
we are forced to give up the antichirality of the superfields $\bar{\hat\La}$
and $\bar{\hat\Phi}$. Contrariwise, if the deformation parameter is 
antiselfdual we are forced to give up the chirality of the superfields 
$\hat\La$ and $\hat\Phi$.   

A nonchiral Seiberg-Witten map for the field $\bar\La^{'}$ is
\beq \label{eq7.10}
\Labpv = \frac{1}{16} \Paabb_{SD} (\daad \bar\La) (\Db\Dbb V) \, . 
\eeq  
It is easy to guess the Seiberg-Witten map for the field $\bar\Phi^{'}$ and 
the gauge superfield $V^{'}$ which correspond to solutions (\ref{eq7.1}) and 
(\ref{eq7.10}). They are 
\beqa 
\Fibpv &=&\quad\!\! \frac{1}{16}\Paabb (\daad \bar\Phi) (\Db\Dbb V) \, ,
\label{eq7.11} \\
V^{'}(V) &=& -\frac{1}{16} \Paabb_{SD} (\daad V) (\Dbb\Db V) + \frac{i}{16} 
P^{\alpha\beta}_{SD} \Da V W_{\beta} \, .  \label{eq7.12}
\eeqa
In the Wess-Zumino gauge the field $\Phi^{'}$ has the usual component 
expansion (\ref{eq2.21}), $\La^{'}$ as given in (\ref{eq5.30}) and $V^{'}$ 
as follows:
\beqa \label{eq7.13}
V^{'} (x,\te,\teb) &=& -i\teb\bar\chi'(x) + \teb\teb \bar M^{'} - \te\sm\teb 
\vm' (x) + i \te\te\teb\lab' (x)   \nonumber \\   
&& - i \teb\teb\te \left[ \la' (x) - \frac{i}{2} \sm \pam\bar\chi'(x) \right] 
+  \frac{1}{2}  \te\te\teb\teb d'(x) \, .
\eeqa
The Seiberg Witten maps for the component fields are
\beqa
\al' (\al,v) &=& - \frac{1}{4} \, \tij_{SD} \pai\al\vj\, ,\label{eq7.14} \\
\eta' (\al,\lab) &=& - \frac{i}{4} \, \tij_{SD} \si \bar \la \paj\al\, ,
\label{eq7.15}  \\ \nonumber \\
A^{'} (A,\psi,v,\la) &=&  - \frac{1}{4} \, \tij_{SD} \left( \pai A \vj  - 
\frac{1}{2\qw} \psi\sij\la \right) , \label{eq7.16} \\ 
\psi{'} (A,\psi,F,v,\la,\lab,d) &=& - \frac{1}{4} \, \tij_{SD} \bigg( \pai\psi
\vj + \frac{i}{\qw}\,\sj\lab\pai A -  \frac{1}{4}\,\sm\sbj\psi f_{im}^{SD} 
\nonumber \\ 
&&\qquad\quad + \frac{i}{4}\,\sij\psi d + \frac{1}{2\qw}\,\sij\la F \bigg)  ,
\label{eq7.17}  \\
F^{'} (\psi,F,v,\lab) &=& - \frac{1}{4} \tij_{SD} \, \left( \pai F \vj - 
\frac{i}{\qw} \, \pai\left( \psi\sj\lab\right) - \frac{1}{2}\, Ff_{ij}^{SD}
\right)  , \label{eq7.18} \\ \nonumber \\
\bar\chi^{'} (v,\la) &=& \,\,\,\, \frac{i}{8} \, \tij_{SD} \sbj\la\vi \, ,
\label{eq7.19} \\ 
\bar M^{'} (\la,\lab) &=& - \frac{i}{16} \, \tij_{SD} \la\sij\lab \, , 
\label{eq7.20}  \\
\vm^{'} (v,\la,\lab) &=& \,\,\,\, \frac{1}{8} \, \tij_{SD} \bigg( \vi \left( 
2\paj\vm -  f_{jm}^{SD} + i\eta_{jm} d\right) + i\eta_{im}\la\sj\lab \!\bigg) ,
\qquad \label{eq7.21} \\
\lab^{'} (v,\lab) &=& - \frac{1}{4} \, \tij_{SD} \bigg( \pai\lab\vj - 
\frac{1}{4} \, \lab f_{ij}^{SD} - \frac{1}{2} \, \sbm\sj\lab\pai\vm \nonumber
\\ &&\qquad\qquad - \frac{1}{2} \, \sbj\sm\pam\lab\vi  \bigg) 
   , \label{eq7.22} \\
\la^{'} (\la,v) &=& - \frac{1}{4} \, \tij_{SD} \bigg( \pai\la\vj - \frac{1}{4}
\, \sm\sbj\la \left( f_{im}^{SD} + 2\pai\vm\right) \nonumber \\
&&\qquad\qquad +  \frac{i}{2} \, \sij\la d 
\bigg)  , 
\label{eq7.23} \\
d^{'} (v,\la,\lab,d) &=& - \frac{1}{4} \, \tij_{SD} \,\bigg( \frac{1}{2} \,
\pai(\vj d) - \pai v^{m}\, {}^{\star}\!f_{jm} + \frac{i}{2} \, \paum\left( \vi 
f_{jm}^{SD} \right)    \nonumber       \\
&&\qquad\qquad + 2\la\sj\pai\lab -  \frac{1}{2} \, \pai(\la\sj\lab) \bigg)  . 
\label{eq7.24}
\eeqa
Taking the complex conjugate of the equations (\ref{eq5.12}) and 
(\ref{eq5.13}), we get the Seiberg-Witten maps for the component fields of 
$\bar\La^{'}$ and $\bar\Phi^{'}$. Note that by complex conjugation on 
Minkowski space $f_{jm}^{SD}$ changes to $f_{jm}^{ASD}$. 

From the component expansion of the field $\bar\La^{'}$ and equation 
(\ref{eq7.21}) it is evident that this solution has the same peculiar 
properties as the nonchiral solution on Minkowski superspace (see section 
\ref{sec:nonchiral}).
    
The determination of the Seiberg-Witten maps on the canonically deformed 
Euclidean superspace with an antiselfdual noncommutativity parameter is 
straight forward and will not be considered further.

\section{Conclusion}

On canonically deformed $N\!=\!1$, $d\!=\!4$ Minkowski superspace there is no 
Seiberg-Witten map for (anti)chiral superfields which is at the same time
(anti)chiral, local and supersymmetric. We have presented Seiberg-Witten maps 
for superfields in each case where one of this requirements is given up and 
constructed the $U(1)$ gauge theory coupled to matter in terms of classical 
component fields using the nonsupersymmetric Seiberg-Witten map. 

The reason such a no-go theorem exists is that the chiral constraint contains 
derivatives, and therefore, in order to keep the constraint for more generic 
terms that can be generated by the Seiberg-Witten map, one needs a nonlocal 
projector. 

On canonically deformed $N\!=\!1$, $d\!=\!4$ Euclidean superspace we face the 
same incompatibility for a general noncommutativity parameter. However, if we 
take the noncommutativity parameter to be selfdual (antiselfdual) there 
exists a chiral (antichiral), local and supersymmetric Seiberg-Witten map for 
the chiral (antichiral) superfields and only antichirality (chirality) is 
broken. Thus it is not possible to construct in terms of the Seiberg-Witten 
map a local and supersymmetric gauge theory with both chiral and antichiral 
superfields on these spaces.

\section*{Acknowledgements}

First and foremost I want to thank Paolo Aschieri and Branislav Jurco for 
collaboration on several aspects of this paper. I also wish to thank 
F. Bachmaier,  W. Behr, C. P. Martin, I. Sachs and J. Wess for useful 
discussions. Furthermore I am indebted to J. Wess for drawing my attention to 
this subject. Last but not least I thank P. Schupp for useful suggestions in 
the initial stage of this project.


\bibliographystyle{diss}
\bibliography{referenzen}

\end{document}